\documentclass[12pt,epsf]{article}
\usepackage{amssymb,amsmath}
\usepackage{graphicx}
\usepackage{caption}
\usepackage{amsthm}
\usepackage{epsfig}
\usepackage{times}
\usepackage[dvips]{color}

\newcommand{\be}{\begin{equation}}
\newcommand{\ee}{\end{equation}}
\newcommand{\bea}{\begin{eqnarray}}
\newcommand{\eea}{\end{eqnarray}}
\newcommand{\beas}{\begin{eqnarray*}}
\newcommand{\eeas}{\end{eqnarray*}}
\newcommand{\ba}{\begin{array}}
\newcommand{\ea}{\end{array}}

\newcommand{\nn}{\nonumber}
\newcommand{\p}{\partial}

\newcommand{\tr}{\mathrm{Tr}}

\newcommand{\nbox}{{\,\lower0.9pt\vbox{\hrule \hbox{\vrule height 0.2 cm \hskip 0.19 cm \vrule height 0.2 cm}\hrule}\,}}

\def\href#1#2{#2}

\begin{document}
\begin{titlepage}
\hfill
\vbox{
    \halign{#\hfil         \cr
           } 
      }  
\vspace*{20mm}
\begin{center}
{\Large \bf Energy trapping from Hagedorn densities of states}

\vspace*{15mm}
\vspace*{1mm}
Connor Behan${}^a$, Klaus Larjo${}^b$, Nima Lashkari${}^a$, Brian Swingle${}^c$, Mark Van Raamsdonk${}^a$

\vspace*{1cm}
{\it ${}^a$
{Department of Physics and Astronomy,University of British Columbia\\
6224 Agricultural Road,Vancouver, B.C., V6T 1W9, Canada} \\ ${}$ \\
${}^b$
{Department of Physics, Brown University, Providence, RI, 02912, USA} \\ ${}$ \\
${}^c$
{Department of Physics, Harvard University, Cambridge MA 02138, USA}
}

\vspace*{1cm}
\end{center}

\begin{abstract}

In this note, we construct simple stochastic toy models for holographic gauge theories in which distributions of energy on a collection of sites evolve by a master equation with some specified transition rates. We build in only energy conservation, locality, and the standard thermodynamic requirement that all states with a given energy are equally likely in equilibrium. In these models, we investigate the qualitative behavior of the dynamics of the energy distributions for different choices of the density of states for the individual sites. For typical field theory densities of states ($\log(\rho(E)) \sim E^{\alpha<1}$), the model gives diffusive behavior in which initially localized distributions of energy spread out relatively quickly. For large N gauge theories with gravitational duals, the density of states for a finite volume of field theory degrees of freedom typically includes a Hagedorn regime ($\log(\rho(E)) \sim E$). We find that this gives rise to a trapping of energy in subsets of degrees of
freedom for parametrically long time scales before the energy leaks away. We speculate that this Hagedorn trapping may be part of a holographic explanation for long-lived gravitational bound states (black holes) in gravitational theories.

\end{abstract}

\end{titlepage}

\vskip 1cm


\section{Introduction}

According to the AdS/CFT correspondence in string theory \cite{malda,agmoo}, gravitational dynamics can arise as an emergent phenomenon in the physics of strongly-coupled large $N$ gauge theories. While there is now a great deal of evidence for the validity of the correspondence, the reason for the emergence of a dual spacetime and its associated gravitational physics is still not well understood.

An intriguing aspect of the correspondence is that many different field theories in different numbers of dimensions are believed to have gravity duals. The physics of these field theories can be quite different, and details of the dual gravitational theories vary, but a very large class of them include Einstein gravity as part of the dual spacetime physics. Thus, Einstein gravity (including the physics of black holes) seems to emerge as a universal sector in the physics of a large class of strongly-coupled large $N$ field theories.

A possible explanation for this universal emergence of gravity is that it is not directly associated with any detailed dynamical properties of the field theory, but rather to ``thermodynamic'' effects, associated with the very large number of degrees of freedom present for large $N$ theories and the large energies (of order $N^2$) required for any non-trivial bulk gravitational dynamics.\footnote{Many connections between gravity and thermodynamics have been made in the past, both within the AdS/CFT correspondence and independent of it. In particular, the physics of black holes is governed by laws that are in perfect analogy with the laws of thermodynamics. A very incomplete list of references is \cite{Bekenstein:1973ur, Bardeen:1973gs, Hawking:1974sw, Jacobson:1995ab, Padmanabhan:2009vy, Verlinde:2010hp}.} In this picture, the dual spacetime metric is thought of as a set of macroscopic thermodynamic functions associated with the field theory (similar to energy density or pressure), and the large
number of degrees of freedom guarantees that the future
evolution of these macroscopic variables (for typical states) is completely determined by the present values through a set of simple macroscopic equations, without further knowledge about the detailed microscopic state.

Furthermore, the microscopic physics enters into the macroscopic equations only in a limited way; for example, in the heat equation, the microscopic physics determines the thermal conductivity as a function of temperature. Once this single function is determined from the underlying microphysics, the dynamics of heat flow in the system is determined. Similarly, it may be that relatively few details of the microscopic physics of large $N$ gauge theories are enough to determine the macroscopic equations that govern the evolution of the metric in the dual spacetime.

Motivated by these observations, we construct in this paper some very crude toy models for large $N$ gauge theories, and ask whether we can reproduce some of the physics associated with the existence of a dual gravitational description by building in only a few key features of the original gauge theory. In particular, we consider stochastic models in which energy is distributed between degrees of freedom living at various sites, and allow these configurations to evolve randomly, demanding only that energy transfers are restricted to neighboring pairs of sites.  Our model is classical and stochastic, but we emphasize that it can still incorporate quantum effects through the choice of the density of states and the transition rates.  By analogy with ordinary thermodynamics and hydrodynamics, we expect that by focusing on just the dynamics of the energy, a classical stochastic picture should suffice to describe the physics.  In essence we are integrating out all degrees of freedom besides the energy density and studying the reduced stochastic dynamics; quantum effects enter indirectly through plausible choices for transition rates and the density of states.

We find that building in only a single feature related to large $N$ gauge theories -- a sufficiently rapid growth in the density of states for the degrees of freedom at a site -- leads to a trapping of energy in localized regions for parametrically long times. These results suggest a direct connection between the Hagedorn (or super-Hagedorn) density of states found for certain subsets of degrees of freedom in large $N$ gauge theories and the long lifetime of gravitational bound states (black holes).\footnote{For a Hagedorn density of states, the density of states grows exponentially with energy as $\rho(E) \propto \exp(\beta_H E)$.} The Hagedorn density of states is also famous as the density of states associated with perturbative strings. Thus, we expect that any theory for which weakly coupled strings play a dominant role in some regime must have a Hagedorn density of states for some subset of its degrees of freedom in some energy regime. Our toy models suggest that these theories will be able to trap energy for long time scales in subsets of the degrees of freedom, so from this perspective, it is natural that string theories are able to describe black holes.

More generally, we find that for slower-than-Hagedorn growth, energy diffuses; small fluctuations about a uniform energy distribution are governed by an ordinary heat equation. For faster-than-Hagedorn growth (found for a certain range of energy in theories whose dual gravitational theory includes small black holes), a clustering behavior is observed; for example, small fluctuations about a uniform energy distribution are governed by the time-reversed heat equation. This clustering behavior for super-Hagedorn growth in the density of states may be part of a field theory explanation for the formation of small black holes in the dual theory.

In gauge theories whose gravity dual admits long-lived evaporating black hole states (e.g. in confining gauge theories where a bulk infrared wall prevents small black holes from falling through a Poincare horizon), the density of states for a localized subset of field theory degrees of freedom typically includes several regimes, with slower than Hagedorn growth at the smallest and largest energies and an intermediate regime characterized by Hagedorn or faster-than-Hagedorn growth. Inserting such a density of states into our stochastic toy model, we find that a large amount of energy localized to a small region will initially spread out to some region of a characteristic size determined by the energy and then remain there for a long time, very slowly losing energy to the surrounding degrees of freedom. Thus, we find behavior qualitatively similar to the formation and evaporation of a black hole.

A physical picture of our results is that the energy is getting trapped within some set of complicated local degrees of freedom that can only be accessed at sufficiently high energy. At these energies local internal rearrangements of the degrees of freedom are much more likely than large scale spatial rearrangements and diffusion is inhibited.  While ordinary conformal field theories and confining gauge theories may not have such a set of many local degrees of freedom, such a picture does emerge from large $N$ confining gauge theories as we discuss in more detail below.  Roughly speaking, once the energy density is above a value set by the confinement scale the system may access a large $N^2$ number of local degrees of freedom and the phenomenon of getting trapped in Hilbert space can occur.

The connection between black holes and Hagedorn or super-Hagedorn densities of states is certainly not new. It is well known that the number of black hole states grows very rapidly with energy, and it has been suggested that this rapid growth may be directly related to the Hagedorn growth of perturbative string states \cite{Susskind:1993ws}. In the AdS/CFT context, the appearance of Hagedorn densities of states in field theory has been linked to the existence of a deconfinement phase transition as temperature is increased; this deconfined phase of the field theory corresponds to large black holes or black branes in the dual gravitational theory (see, for example \cite{Witten:1998zw, Sundborg:1999ue,Aharony:2003sx}). The novel aspect of the present work is that we relate the behavior of the density of states to the {\it dynamics} of energy transfer between degrees of freedom in the field theory. We are thus able to suggest a mechanism by which energy can be trapped for long periods of time in localized regions of an interacting, translationally-invariant open system, providing a possible field theory explanation for the existence of long-lived (but unstable) black hole states in the dual theory.

The paper is organized as follows. In section 2, we motivate and describe the basic stochastic models
that we consider. We show that Hagedorn densities of states are naturally associated with static or
nearly static energy distributions. In section 3, we derive differential equations that govern the
evolution of the energy distribution in the thermodynamic and continuum limits of our model,
and investigate how the growth rate in the density of states affects the evolution.
In section 4, we review relevant results on the density of states for field theory models with a
gravity dual. In section 5, we argue that using the density of states for a typical confining
gauge theory with gravity dual in our models generically results in behavior qualitatively similar to the
expected field theory description of an evaporating black hole, even though our model incorporates
very few features of a realistic field theory. We consider a specific model
 and employ PDE techniques and numerics to verify our qualitative expectations for the evolution of energy distributions within the model.

\section{Basic setup}

At the most basic level, a field theory is a collection of degrees of freedom arranged in a translationally-invariant way in some number of dimensions. These degrees of freedom can be excited to different configurations depending on how much energy is available to them. As a simple model of this, we consider a lattice of sites, with some unspecified degrees of freedom on each site that can exist in $\rho(n)$ possible configurations given energy $n$ (which we take to be integer-valued for now). The configuration of the model at a given time is specified by the collection $\{(n_r, k_r)\}$ where $n_r$ is number of units of energy residing on site $r$, and $k_r \le \rho(n_r)$ is a positive integer specifying the state of the degrees of freedom on site $r$. To incorporate dynamics into the model, we need to specify some transition rates $\{(n_r, k_r)_t\} \to \{(n_r, k_r)_{t + dt}\}$. We constrain these by locality, conservation of energy and the principle of detailed balance, but otherwise simply postulate that
all transitions consistent with these basic constraints are equally likely. Our construction gives rise to a particular set of stochastic equations for the evolution of a probability distribution of $p(\{n_r\})$ over the possible energy distributions in our model. We find that the most general evolution equations satisfying our constraints depend on only two detailed features of the underlying physics: the density of states $\rho(E)$ for each site, and another function $C(E)$ related to the energy-dependence of transition rates.

\subsection{Master equations for evolution}

To define a precise stochastic model using our basic setup, we assign a probability $p(\{n_r\})$ to a particular energy distribution $\{n_r\}$. We do not keep track of the detailed state, labelled by $k_r$, of each site.  The time evolution of such an ensemble is governed by the transition rates $W_{(\{n_r\} \to \{n'_r\})}$, giving the probability per unit time that the system in configuration $\{n_r\}$ will make a transition to configuration $\{n'_r\}$. The time evolution is governed by a master equation\footnote{For a general review of stochastic processes and master equations, see for example \cite{vankampen}.}
\[
\partial_t p(\{n_r\}) = \sum_{\{n'_r\}} \left[ p(\{n'_r\}) W_{\{n'_r\} \to \{n_r\}} - p(\{n_r\}) W_{\{n_r\} \to \{n'_r\}} \right] \; .
\]
For our system, we assume energy conservation and nearest neighbor interactions, so that the rate vanishes unless the total energy is conserved and the final state differs from the initial state only at a neighboring pair of sites. We assume that locality in the interactions further implies that the rate for a transition $\{\dots,n_a,n_b,\dots\} \to \{\dots,n'_a,n'_b,\dots\}$ involving neighboring sites $a$ and $b$ does not depend on the state of the degrees of freedom on the other sites. Finally, we assume a discrete translation/rotational invariance such that this rate $W_{(n_a,n_b) \to (n'_a,n'_b)}$ depends only on $n_a, n_b, n'_a$, and $n'_b$, and not on the specific sites $a$ and $b$ involved. With these restrictions, the master equation takes the form
\be
\label{mast}
\partial_t p(\{n_r\}) = \sum_{\left < a, b \right >} \sum_{k \ne 0} \left[ p(\{n_r |n_a + k, n_b - k\}) W_{(n_a + k ,n_b - k) \to ( n_a,n_b)} - p(\{n_r\}) W_{(n_a,n_b) \to (n_a + k ,n_b - k)} \right] \; ,
\ee
where $\langle a,b \rangle$ indicates nearest-neighbor sites and $p(\{n_r| n_a + k, n_b - k\})$ is the probability for the distribution that is obtained from $\{n_r\}$ by adding $k$ units of energy to site $a$ and subtracting $k$ units of energy from site $b$.
From this, we can give a general result for the evolution of the energy expectation value at a particular site
\[
\bar{n}_c \equiv \sum_{\{n_r\}} p(\{n_r\}) n_c\; .
\]
We find
\be
\label{macro}
\partial_t \bar{n}_c = \left\langle \sum_{\left < c, a \right >} \sum_{k \ne 0} k \; W_{(n_c,n_a) \to (n_c + k ,n_a - k)} \right\rangle\; ,
\ee
where the first sum is over sites $a$ that are nearest neighbors to $c$.

\subsubsection*{Equilibrium and detailed balance}

The master equations above act independently in each sector with a given total energy. In equilibrium, a basic assumption of thermodynamics is that all states with a given energy are equally likely. If $\rho(n)$ is the number of states on a given site with energy $n$, then the number of states with the distribution $\{n_r\}$ is $\prod_r \rho(n_r)$, so we should have for an equilibrium configuration
\be
\label{Peq}
p(\{n_r\}) = p_0 \prod_r \rho(n_r)
\ee
where $p_0$ is a constant that can depend on the total energy. For such a configuration, the right-hand side of (\ref{mast}) must vanish, but we should generally expect a stronger condition, the Principle of Detailed Balance. This requires that in a given time, the number of transitions in the ensemble from distributions $\{n_r\}$ to $\{n'_r\}$ should equal the number of transitions from $\{n'_r\}$ to $\{n_r\}$. This is equivalent to saying that for the equilibrium configuration (\ref{Peq}), each term in the square brackets in (\ref{mast}) should vanish separately, i.e.
\[
p(\{n_r | n_a + k, n_b - k\}) W_{(n_a + k ,n_b - k) \to ( n_a,n_b)} =  p(\{n_r\}) W_{(n_a,n_b) \to (n_a + k ,n_b - k)} \; .
\]
Using (\ref{Peq}), this gives a condition on the transition rates, that
\be
\label{Wconst}
\rho(n_a + k) \rho(n_b - k) W_{(n_a + k ,n_b - k) \to ( n_a,n_b)} = \rho(n_a) \rho(n_b) W_{(n_a,n_b) \to (n_a + k ,n_b - k)} \; .
\ee

\subsubsection*{Transition rate democracy}

Since we are interested in effects that do not depend strongly on the underlying microscopic dynamics, we will simply assume that the transition rate from a given initial energy distribution $(n_a, n_b)$ of a pair of sites to various possible final distributions $(n_a + k, n_b - k)$ is proportional to the number of available final states with that distribution. This gives:
\be
\label{Wconst2}
W_{(n_a,n_b) \to (n_a + k ,n_b - k)}  =  C(n_a,n_b) \rho(n_a + k) \rho(n_b - k) f(k)   \; ,
\ee
where we allow a function $f(k)$ (taken to be even) that can be used to introduce an additional dependence on the amount of energy transferred. As an example, we might want to allow only transitions up to a certain maximum energy transfer, using a function
\[
f(k) = \theta(E_{\mathrm{max}} - |k|)
\]
Combining this with the constraint (\ref{Wconst}) above, we find that
\[
C(n_a,n_b) = C(n_a + k,n_b - k) \qquad \qquad f(k) \ne 0
\]
but using this relation repeatedly gives\footnote{Here, we are assuming that any configurations with the same energy can be connected by a series of allowed transitions i.e. that there are no superselection sectors.}
\[
C(n_a,n_b) = C((n_a + n_b)/2) \; ,
\]
where we have included the factor of $1/2$ for later convenience.
Thus, we have
\be
\label{generalW}
W_{(n_a,n_b) \to (n_a + k ,n_b - k)}  = C((n_a + n_b)/2) \rho(n_a + k) \rho(n_b - k) f(k) \; .
\ee

\subsubsection*{Choices for $W$}

We now comment on a few specific choices for transition rates of the form (\ref{generalW}). The simplest choice is
\[
W_{(n_a,n_b) \to (n_a + k ,n_b - k)} = C \rho(n_a + k) \rho(n_b - k) \; .
\]
This model arises if we simply assume that all microscopic transition rates between the individual quantum states states are equal.

We can also restrict the number of units of energy transferred in a step to obtain
\be
\label{model1}
W_{(n_a,n_b) \to (n_a + k ,n_b - k)} = C \theta(E_{\mathrm{max}} - |k|) \rho(n_a + k) \rho(n_b - k) \; .
\ee

For another physically motivated solution, consider the situation where nearest neighbor pairs interact with some rate $q$ (independent of the pair and the energy on the sites) and where in each interaction, the two members of the pair instantly thermalize. In this case, the state after the interaction will be $(n'_a,n'_b)$ with probability proportional to $\rho(n'_a) \rho(n'_b)$ with the probabilities summing to one. This gives a transition rate
\[
W_{(n_a,n_b) \to (n_a + k ,n_b - k)} = C {\rho(n_a + k) \rho(n_b - k) \over \sum_l \rho(n_a + l) \rho(n_b - l)}
\]
which is also of the form (\ref{generalW}). With a constraint on the number of units of energy transferred, this gives
\be
W_{(n_a,n_b) \to (n_a + k ,n_b - k)} = C  \theta(E_{\mathrm{max}} - |k|) {\rho(n_a + k) \rho(n_b - k) \over \sum_l \rho(n_a + l) \rho(n_b - l)} \; .
\label{model2a}
\ee
Physically, the denominators in these expressions arise because we have introduced two time-scales into the problem; the final rate is the product of a basic reaction rate with a probability for various outcomes of the reaction (because the interaction that determines the final state happens very fast such that the probability for changing to a different state assuming the pairs interact is not much smaller than one).

We can obtain a simpler model that should have similar qualitative behavior to (\ref{model2a}) by replacing $\sum_l \rho(n_a + l) \rho(n_b - l)$  with  $\rho^2((n_a + n_b)/2)$. This gives
\bea
\label{model2}
W_{(n_a,n_b) \to (n_a + k ,n_b - k)} &=& C  \theta(E_{\mathrm{max}} - |k|) {\rho(n_a + k) \rho(n_b - k) \over \rho^2((n_a + n_b)/2)} \; .
\eea

\subsection{Static distributions for Hagedorn density}

We would now like to analyze the physics of the class of master equations derived in the previous section. Starting from the general form (\ref{generalW}) of the transition rate, the equation (\ref{macro}) for the evolution of the average energy at a site becomes:
\be
\partial_t \bar{n}_c = \left\langle \sum_{\left < c, a \right >} \sum_{k \ne 0} k \; C((n_c + n_a)/2) \rho(n_c + k) \rho(n_a - k) f(k) \right\rangle\; .
\ee
In the next section, we will study this equation in the thermodynamic limit where fluctuations are small and derive a differential equation for the evolution of the energy distribution in a continuum limit where variations are small on the scale of the lattice spacing. However, we note that even at this stage, a special role is played by the Hagedorn density of states $\rho(n) = B e^{A n}$. With this choice, we find
\beas
\partial_t \bar{n}_c &=& \left\langle \sum_{\left < c, a \right >} \sum_{k \ne 0} k \; C((n_c + n_a)/2) ( B e^{A(n_c + k)})  (B e^{A(n_a - k)}) f(k) \right\rangle \cr
&=& \left\langle \sum_{\left < c, a \right >}  \; C((n_c + n_a)/2) B^2 e^{A(n_c + n_a)} \sum_{k \ne 0} k f(k) \right\rangle \cr
&=& 0 \; ,
\eeas
where in the last line we have used the fact that $f$ is an even function of $k$. Thus, for an exact Hagedorn density of states, the average energy at each site is constant. This is simple to understand: if we have two neighboring sites, the number of states with energy $n_a$ on one site and $n_b$ on the other depends only on $n_a + n_b$ for a Hagedorn density, so by our assumptions above, it is equally likely for a unit of energy to move from $a$ to $b$ as it is for the energy to move from $b$ to $a$.

For the result we have just derived to hold exactly, the Hagedorn density must hold even for negative energies; more realistically, we should restrict to positive energy i.e. we should take $\rho(n) = 0$ for $n < 0$. In this case, starting with some distribution for which all sites have nonzero energy, the system will evolve with no change in the average energies until some members of the ensemble have zero energy for one or more sites. Since we cannot transfer any more energy from the sites with zero energy, these sites will gain energy on average, so there will be a slow diffusion of energy to sites for which the ensemble includes states with zero energy on these sites.

\section{Macroscopic equations in the continuum limit}

Returning to an unspecified density of states, we will now analyze the equation (\ref{macro}) for the evolution of the average energy in the model (\ref{mast}) with transition rates of the form (\ref{generalW}). We have
\be
\partial_t \bar{n}_c = \left\langle \sum_{\left < c, a \right >} \sum_{k \ne 0} k \; C((n_c + n_a)/2) \rho(n_c + k) \rho(n_a - k) f(k) \right\rangle\; .
\ee
For the large $N$ theories that we consider (or more general theories where the individual sites represent enough degrees of freedom so that a thermodynamic limit can be assumed), we can replace the expectation value of the function of $n_a$ here with the function of the expectation value to get an evolution equation for distribution of average energies,
\be
\partial_t n_c =  \sum_{\left < c, a \right >} \sum_{k \ne 0} k \; C((n_c + n_a)/2) \rho(n_c + k) \rho(n_a - k) f(k) \; ,
\ee
where now all $n$s represent ensemble averages.

Let us now take a continuum limit of this equation. Specializing to one spatial dimension, we can take $c = x$ and $a \in \{x+\epsilon, x - \epsilon \}$ to obtain
\beas
\partial_t n(x) &=&  \sum_{k \ne 0} k f(k) \; \left\{ C((n(x) + n(x +\epsilon))/2) (\rho(n(x) + k) \rho(n(x +\epsilon) - k) \right. \cr && \qquad \qquad +  \; \left. C((n(x) + n(x -\epsilon))/2) (\rho(n(x) + k) \rho(n(x -\epsilon) - k) \right\} \; ,
\eeas
We also rescale energies $k \to \delta k$ so that $k$ becomes continuous in the limit $\delta \to 0$, and assume that $f(k)$ falls off fast enough that only infinitesimal values of $k$ contribute in the limit. In this case, taking the limits $\delta, \epsilon \to 0$ while rescaling $C$ to leave a finite result gives
\[
\partial_t E(x) = \partial_x \left\{ - \rho^2(E(x)) C(E(x)) \partial_x {\textup{d} \ln(\rho) \over \textup{d}E} \right\}
\]
where we have denoted the continuum energy distribution by $E(x)$. Generalizing to higher dimensions, we find
\[
\partial_t E(x) = \nabla \cdot \left\{ - \rho^2(E(x)) C(E(x)) \nabla {\textup{d} \ln(\rho) \over \textup{d}E} \right\} \; .
\]

It is interesting that the right side vanishes identically if and only if we have a Hagedorn density of states $\rho(E) \propto e^{A E}$.

The expression $\ln(\rho(E))$ is (up to a constant) the microcanonical definition of entropy, so we have
\[
{\textup{d} \over \textup{d}E} \ln(\rho(E)) = {\textup{d}S \over \textup{d}E} = {1 \over T} = \beta \; .
\]
In the end we have simply
\be
\label{general}
\partial_t E(x) = \nabla \cdot \left\{ - \rho^2(E(x)) C(E(x)) \nabla \beta \right\} \; .
\ee
For the simple model (\ref{model1}), we obtain
\be
\label{Eone}
\partial_t E(x) = C \nabla \cdot \left\{ - \rho^2(E(x)) \nabla \beta(E(x))  \right\} \; .
\ee
while for the model (\ref{model2}), we get simply
\be
\partial_t E(x) = - C \nabla^2 \beta(E(x))   \; .
\label{Etwo}
\ee
We note that by defining $\tilde{\beta}$ with the property that
\[
\tilde{\beta}'(E) = \rho^2(E(x)) C(E(x)) \beta'(E)
\]
the general model (\ref{general}) reduces to
\be
\partial_t E(x) = - \nabla^2 \tilde{\beta}(E(x))   \; ,
\label{EtwoA}
\ee
the same form as the simplest model (\ref{Etwo}). Since $\rho$ and $C$ are positive functions, $\tilde{\beta}'(E)$ has the same sign as $\beta'(E)$ for any $E$. We will see that the qualitative dynamics in the model is largely determined by the sign of $\tilde{\beta}'(E)$ as a function of $E$, so studying the simplest model with $\tilde{\beta} =  \beta$ should give us a good idea of the general behavior.

\subsubsection*{Perturbations around a uniform density}

As a start towards understanding the qualitative behavior of our macroscopic equation, consider small perturbations about a uniform energy distribution (which is a time-independent configuration for any density of states).  Expanding (\ref{general}) around a uniform density $E_0$, we find that the equation governing perturbations is
\[
\partial_t e = - \rho^2(E_0) C(E_0) {\textup{d}^2 \ln(\rho(e)) \over \textup{d} e^2}(E_0) \nabla^2 e,
\]
where $e=E-E_0$. Thus, we get exactly the heat equation or the inverse heat equation depending on whether $\textup{d}^2 \ln(\rho) / \textup{d} E^2 = d \beta / d E$ is negative or positive, with no time dependence for the Hagedorn density. In summary, we have the following possibilities:
\begin{itemize}
\item
$\textup{d} \beta / \textup{d} E < 0$: diffusive behavior, inhomogeneities decrease over time
\item
$\textup{d} \beta / \textup{d} E > 0$: unstable clustering of energy, inhomogeneities increase over time
\item
$\textup{d} \beta / \textup{d} E = 0$: static energy distribution
\end{itemize}
In realistic field theories, the density of states may be a function of energy that exhibits more than one of these behaviors. In this case, we can have different qualitative behaviors for the evolution depending on the local energy density.

\section{Density of states in large $N$ field theories}

We have seen that Hagedorn densities of states play a special role in our simple models, leading to static energy distributions, or equivalently a trapping of energy in certain subsets of degrees of freedom. This behavior is reminiscent of the expected behavior of black holes in gravitational theories. There, large amounts of energy sent in from an asymptotic region can form a black hole; the energy is trapped in a localized region (presumably described by a subset of the degrees of freedom in the fundamental description) for a time scale that goes to infinity in the classical limit. In this section, we recall that Hagedorn densities of states appear naturally in large $N$ gauge theories, some of which provide non-perturbative descriptions of quantum gravitational theories via the AdS/CFT correspondence. Thus, the Hagedorn trapping of energy in these theories may be part of the explanation for the existence of black hole states in the dual gravitational theories.

Consider first a simple model with two matrices $X$ and $Y$ of degrees of freedom, with each matrix element describing a harmonic oscillator \cite{Aharony:2003sx}. We assume no interactions, but demand that the states are invariant under gauge transformations that act on the matrices as $X \to U X U^{-1}$ and $Y \to U Y U^{-1}$. If we define matrix creation operators $A_X^\dagger$ and $A_Y^\dagger$ in the usual way, then the gauge-invariant states are constructed with traces of products of these operators acting on a Fock space vacuum state. If each oscillator carries a unit of energy, then the number of single-trace states with energy $E$ is on the order of $2^E = e^{E \ln(2)}$, since we must have $N$ operators in the trace and each can be $A_X^\dagger$ or $A_Y^\dagger$. Some of these states are equivalent by cyclicity of the trace, but the overcounting is by at most a factor of E, so the leading asymptotic behavior of the density of states is Hagedorn. Including multiple trace states does not affect this
conclusion.

As discussed in \cite{Sundborg:1999ue,Aharony:2003sx}, this Hagedorn behavior is obtained in the weak coupling limit of any large $N$ gauge theory with a discrete set of matrix degrees of freedom (e.g. the momentum modes in a finite volume field theory). However, the analysis breaks down for energies of order $N^2$. For traces with of order $N^2$ matrices, algebraic identities allow us to rewrite certain traces in terms of other traces (e.g. $\tr(M^3) = {3 \over 2} \tr(M) \tr(M^2) - {1 \over 2} \tr(M)^3$ for a $2 \times 2$ matrix) and the result is that the density of states for energies of order $N^2$ crosses over to $e^{E^\alpha}$ where $\alpha < 1$.

Away from weak coupling, the expected behavior of the density of states in large $N$ theories with a discrete spectrum of states (for example, large $N$ CFTs compactified on a sphere or a circle) was discussed in detail in \cite{Aharony:2003sx} (see sections 3.1, 6.5, and 7.2) and in \cite{Aharony:2005bm}, appendix A.2. Here, the behavior may be inferred from the gravitational dual description.
Typically we have $\ln(\rho(E)) \sim E^\alpha$ with $\alpha < 1$ for energies of order 1 (corresponding to exciting perturbative gravity / field theory modes in the bulk), then $\alpha=1$ (Hagedorn behavior) for energies of order $\lambda^{1 \over 4}$ where $\lambda$ is the t'Hooft coupling (corresponding to exciting string excitations in the bulk). At energies of order $N^2$, we first encounter a regime with $\alpha > 1$ (corresponding to small black holes in the bulk) and finally, above a certain energy of order $N^2$, this behavior crosses over to $\alpha = d/(d+1)$ (corresponding to large AdS black holes).

Thus, for large $N$ gauge theories with a mass gap, at both strong and weak coupling, we have a Hagedorn regime in the density of states at intermediate energies, with slower growth, $\rho(E) \sim \exp(E^{\alpha < 1})$, for energy densities of order $N^2$ and for energy densities smaller than order $\lambda^{1 \over 4}$ in strongly coupled theories with gravity duals. In the next section, we will explore the dynamics of energy distributions for such densities of states in our simple model. We focus on spherically symmetric energy distributions which initially have a large density of energy in a small-volume region. We will see that the existence of an intermediate-energy Hagedorn regime in the density of states dramatically increases the time for the energy to leak away to infinity.

This is in accord with our expectations for the gravity dual of such a theory, where a very localized collection of energy should correspond to a bulk configuration that will form a black hole. For gauge theories with a mass gap, the bulk dual has an infrared wall to which the black hole will fall and then very slowly evaporate.

Note that for conformal holographic field theories on $R^{d,1}$, the bulk geometry dual to the vacuum state is the Poincare patch of AdS space. In this geometry, massive objects fall into a horizon in finite proper time that is typically much shorter than the black hole evaporation time. Since only this shorter time scale is visible in the field theory evolution, it should be difficult to deduce the long lifetime of gravitational bound states from the field theory in this conformal case. Here, the dynamics of an energy distribution would be expected to be dominated by a simple spreading behavior that corresponds to the bulk object falling towards the horizon. These expectations are consistent with the fact that the entropy density for a conformal field theory is related to the energy density (according to conformal invariance) by $S \sim E^{d/(d+1)}$. This suggests a density of states $\rho(E) \sim \exp(E^{d/(d+1)})$ that corresponds in our model to a simple diffusive (spreading) behavior.

\section{Evolution of energy distributions}

In this section, we study the evolution of simple energy distributions in our model, choosing the density of states to be similar to those in large $N$ gauge theories with gravity duals that support long-lived evaporating black hole states.

As we have shown above, our most general model (\ref{general}) can be rewritten to take the form
\be
\partial_t E(x) = -  \nabla^2 \tilde{\beta}(E(x))   \; ,
\label{Etwo2}
\ee
where $\tilde{\beta}$ is a function of energy for which $\tilde{\beta}'(E)$ has the same sign as the derivative $\beta(E) = \textup{d}S/\textup{d}E$.

We begin by comparing (\ref{Etwo2}) with the energy conservation equation
\[
\partial_t E = - \nabla \cdot J_E \; .
\]
We see that the physical content of (\ref{Etwo2}) may be summarized by saying that the energy current is
\be
\label{current}
J_E = \nabla ( \tilde{\beta}(E(x)) ) =  \tilde{\beta}'(E(x)) \nabla E(x) \; .
\ee

For typical field theories, $\beta'(E) < 0$, so energy flows from regions of higher energy to regions of lower energy. This is standard diffusive behavior.

In theories with a regime in the density of states where $\beta'(E) > 0$, (\ref{current}) shows that the energy flows from regions of lower energy to regions of higher energy. Thus, with a very rapidly growing density of states, we expect a clustering behavior in which local regions of larger energy draw even more energy from the surrounding regions.\footnote{This can be understood very simply. As an example, consider a situation with energy E on each of two sites. If the number of possible configurations for 2E units of energy on a single site is larger than the square of the number of configurations for E unit of energy on a site, then by our democratic evolution rules, the system is more likely to end up in the configuration with $2E$ units of energy on a single site. The condition $\rho(2E) > \rho(E)^2$ is equivalent to saying that the growth of $\log \rho$ is faster than linear.} As we discussed in the previous section, this behavior in the density of states is found in holographic theories whose dual
gravity description includes small black hole states. Our observations suggest that the $\beta'(E) > 0$ behavior may in fact underlie the dynamics that allows the formation of these small black holes.

For theories with a Hagedorn regime in the density of states, such as the large $N$ theories discussed in the previous section, (\ref{current}) shows that the energy current is zero in regions where the local energy density is in this Hagedorn regime. In this case, dynamics is frozen in the regions of space where the energy density corresponds to Hagedorn behavior. We will see below that the boundaries of such regions can move, so energy is not trapped forever as long as there is a lower-energy regime in the density of states with slower than Hagedorn growth.

Our observations here are consistent with those at the end of section 3 based on the analysis of small perturbations around a uniform energy distribution.

\subsection{Spherically symmetric initial energy distributions}

In this section, we consider the evolution of smooth spherically symmetric energy distributions which are monotonically decreasing in the radial direction.

For a density of states with $\beta'(E) < 0 $ at all energy scales, the energy current points radially outward everywhere, and the energy diffuses away to infinity. If $\tilde{\beta}'(E)$ is bounded above by some negative constant $-C$, then we have
\[
|J_E| = |\tilde{\beta}'(E)||\nabla E| \ge C |\nabla E|
\]
so the magnitude of the energy current is at least as large as for the ordinary diffusion equation with diffusion constant $C$. Thus, an initially localized collection of energy will spread out at least as quickly as a system governed by some diffusion equation. In particular, if a certain amount of energy $E_0$ is contained in a region of size $R_0$, the amount of time before the energy in this region is $\epsilon E_0$ can be bounded above by a time that is independent of $E_0$.  This follows because the diffusion equation is linear, so if $f$ is a solution of the diffusion equation then so is $E_0 f$.  Thus the timescale in $f$, which is set by the spatial profile (and in particular the scale $R_0$) and the diffusion constant, is independent of the overall energy $E_0$.

\begin{figure}
\centering
\includegraphics[width=0.4\textwidth]{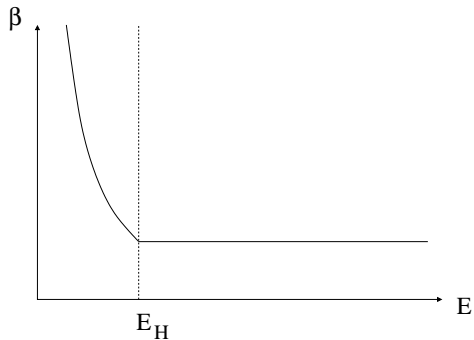}
\caption{Energy dependence of $\beta$ for density of states with a Hagedorn regime for $E>E_{\mathrm{H}}$.}
\label{two_regimes}
\end{figure}

\begin{figure}
\centering
\includegraphics[width=0.7\textwidth]{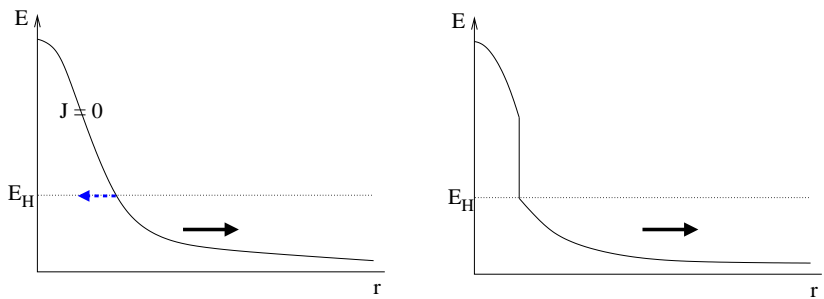}
\caption{Evolution of an energy distribution with a Hagedorn regime in the density of states for $E>E_{\mathrm{H}}$. The interface between Hagedorn and non-Hagedorn energy density moves inward with time, allowing the energy to escape eventually.}
\label{distfig3}
\end{figure}

\subsubsection*{Effects of a high-energy Hagedorn regime}

Now let us consider the effects of a Hagedorn regime in the density of states at energies $E > E_{\mathrm{H}}$, with $\beta'(E) < 0 $ for smaller energies, as depicted in figure (\ref{two_regimes}). In this case, consider an initial energy distribution for which the central energy is larger than $E_{\mathrm{H}}$, as shown in figure \ref{distfig3}. The energy current is strictly zero in the regions with $E > E_{\mathrm{H}}$, while energy flows outward in the surrounding region where $E > E_{\mathrm{H}}$. One might expect that the energy in the Hagedorn region is trapped there forever; however, integrating (\ref{Etwo2}) over a ball with radius R, we find that flux of energy going out of the ball has a finite limit as $R$ approaches the radius $R_{\mathrm{H}}$ where $E = E_{\mathrm{H}}$. Thus, the interface point must move inward with time, as indicated by the dashed arrow in figure \ref{distfig3}. The resulting energy distribution necessarily exhibits a step-function behavior at $R_{\mathrm{H}}$, since the
energy distribution for $R<R_{\mathrm{H}}$ must be exactly the same as the
initial energy distribution.

As an explicit example, we consider in Appendix A the time evolution of a one-dimensional energy distribution that is initially
\[
E(x,0) = \left\{ \ba{ll} E_0 & |x| < a \cr 0 & |x| > a \ea  \right.\; ,
\]
comparing the evolution with constant $\beta'(E) = -C$ (which gives a standard heat equation)  to the evolution with $\beta'(E) = -C$ below energy $E_{\mathrm{H}}$ but $\beta'(E) = 0$ (a Hagedorn regime) above energy $E_{\mathrm{H}}$. We are able to find an exact solution for the evolution in both cases. Comparing the energy in the region $|x|>a$ as a function of time for the two cases, we find that when $E_0 \gg E_{\mathrm{H}}$, the behavior for early times is
\[
E(|x|>a, t) = \left\{ \ba{ll} 2 E_0 \sqrt{t \over \pi} & {\rm no \; Hagedorn} \cr 4 E_{\mathrm{H}} \sqrt{t \over \pi} & {\rm with \; Hagedorn} \ea \right. \; .
\]
Thus, the rate of energy flow away from the region where the energy is initially concentrated is parametrically smaller when the density of states exhibits Hagedorn behavior for energies above a threshold.

\subsubsection*{Effects of a low-energy Hagedorn regime}

\begin{figure}
\centering
\includegraphics[width=0.9\textwidth]{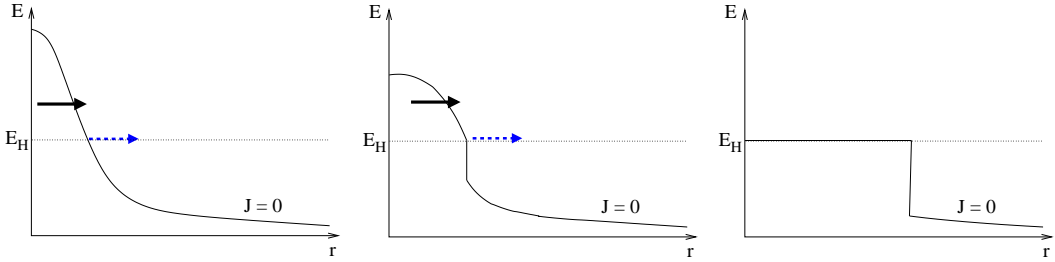}
\caption{Evolution of an energy distribution with a Hagedorn regime in the density of states for $E < E_{\mathrm{H}}$. The interface between Hagedorn and non-Hagedorn energy density moves outward with time, resulting in a flattened distribution with energy density $E_{\mathrm{H}}$.}
\label{lowHag}
\end{figure}

Next, we consider the effects of a Hagedorn regime for low energies $E < E_{\mathrm{H}}$, with $\beta'(E) < 0$ for higher energies. In this case, for an initial energy distribution with central energy larger than $E_{\mathrm{H}}$, the energy current is strictly zero in the regions with $E < E_{\mathrm{H}}$, while energy flows outward in the central region where $E > E_{\mathrm{H}}$. Since the outward flux of energy just inside the radius $R_{\mathrm{H}}$ where $E = E_{\mathrm{H}}$ is finite, the interface point must move outward with time, allowing the central concentration of energy to spread eventually to a uniform disc with energy density $E_{\mathrm{H}}$, as shown in figure \ref{lowHag}. As above, the energy distribution necessarily exhibits a step-function behavior at $R_{\mathrm{H}}$, since the energy distribution for $R>R_{\mathrm{H}}$ must be exactly the same as the
initial energy distribution.

\subsubsection*{Effects of an intermediate-energy Hagedorn regime}

\begin{figure}
\centering
\includegraphics[width=0.9\textwidth]{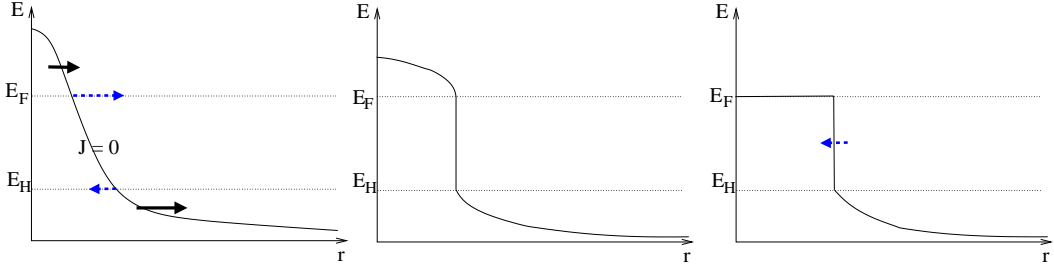}
\caption{Evolution of an energy distribution with a Hagedorn regime in the density of states for $E_{\mathrm{H}} < E < E_{\mathrm{F}}$.}
\label{intHag}
\end{figure}

Finally, we consider the effects of an intermediate-energy Hagedorn regime, with $\beta'(E)=0$ for $E_{\mathrm{H}} < E < E_{\mathrm{F}}$ and $\beta'(E) < 0$ for both high and low energies. This behavior of $\beta$ is similar to what we expect for certain holographic confining gauge theories, as we discussed in section 4. In this situation, we have a combination of the behavior in the previous two cases, as shown in figure \ref{intHag}. The energy spreads outward in the central region and in the outer region, requiring the inner boundary of the Hagedorn region to move outward while the outer boundary of the Hagedorn region moves inward. These eventually come together, leaving a discontinuity at which the energy jumps from $E_{\mathrm{H}}$ to $E_{\mathrm{F}}$. We expect that the central distribution eventually spreads out to one with constant energy density $E_{\mathrm{F}}$, after which it evolves as we discussed in the case of a high-energy Hagedorn regime. To check this intuition, we have performed numerical
simulations of the evolution equation in our model for $\beta(E)$ with this behavior. The results, described in detail in section 5.2, confirm the qualitative picture in figure \ref{intHag}.

\subsubsection*{Comparison with expectations from holographic confining gauge theory}

The behaviour shown in figure \ref{intHag} is qualitatively similar to what we expect for the evolution of a highly concentrated energy distribution in a holographic confining gauge theory. Such an initial energy distribution might arise from a high-energy collision of glueballs. In this case, we expect that the large initial concentration of energy spreads out relatively quickly to a metastable ``plasma ball'' \cite{Aharony:2005bm} with uniform energy density. At this stage, the energy distribution would be similar to the third frame in figure \ref{intHag}. The plasma ball then slowly evaporates, with the boundary of the deconfined plasma region moving inward until the plasma ball has completely evaporated.

 The formation of a plasma ball and its subsequent evaporation corresponds in the dual gravity picture to the formation and evaporation of a black hole sitting at the infrared wall of the dual spacetime. Thus, the metastable black hole in the gravity picture corresponds in our model to the uniform energy distribution in the central region in the third frame of figure \ref{intHag}.
In our toy model, we have seen that the Hagedorn behavior $\beta'(E)=0$ for $E > E_{\mathrm{H}}$ leads to a parametric suppression of the rate of energy loss from such a configuration (at least at early times). Hence, it is plausible that the Hagedorn regime in the density of states for the actual holographic confining gauge theory may be part of the explanation for the long lifetime of the black hole state in dual gravitational theory.

\subsection{Numerical results}

In this section, we present some numerical results for the evolution of spherically-symmetric energy distributions in a model for which the density of states has an intermediate-energy regime with nearly Hagedorn behavior, in order to check the qualitative expectations discussed in the previous section.

\begin{figure}
\centering
\includegraphics[width=.6\textwidth]{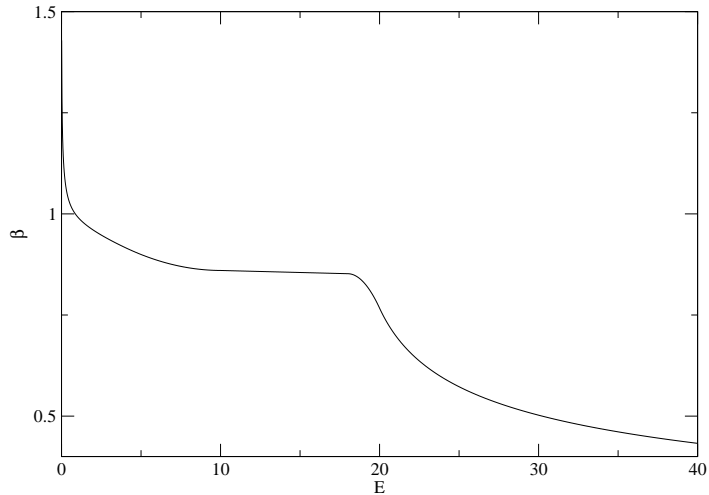}
\caption{An inverse temperature function with $E$ on the horizontal axis and $\beta$ on the vertical one. It shows smooth transitions into and out of the Hagedorn phase and was plotted for $E_{\mathrm{F}} = 20$.}
\label{interpolated-beta}
\end{figure}

It is possible to treat the Cauchy problem for the equation (\ref{Etwo2}) with finite difference schemes. We have used the Crank-Nicolson method, a second-order approximation stable for large timesteps, to demonstrate some of the qualitative features described in this paper
In its most well known form, Crank-Nicolson is used to simulate the heat equation. Since it is an implicit Runge-Kutta method, it requires a system of linear algebraic equations to be solved at each time. Applying the Crank-Nicolson method to our equation results in a set of non-linear algebraic equations which we solve by Newton's method.
Figure \ref{interpolated-beta} shows the $\beta(E)$ function used in these simulations. Motivated by the behavior for holographic confining gauge theories, we choose $\beta$ to decay with a power of $-\frac{1}{10}$ at low energies and with a power of $-\frac{1}{4}$ for high energies $E > E_{\mathrm{F}}$ with a nearly flat region in between. We do not include a regime with $\beta' > 0$, since this complicates the numerics significantly.

Most of the initial conditions that were tested had the form
\[
E(x,0) = 20 E_{\mathrm{F}} \left( \frac{1}{1 + x^2} \right)^{\frac{4}{3}} \; ,
\]
chosen to give a smooth, localized energy distribution with central density significantly greater than the value $E_F$ corresponding to the crossover to the high-energy regime in the density of states.

\begin{figure}
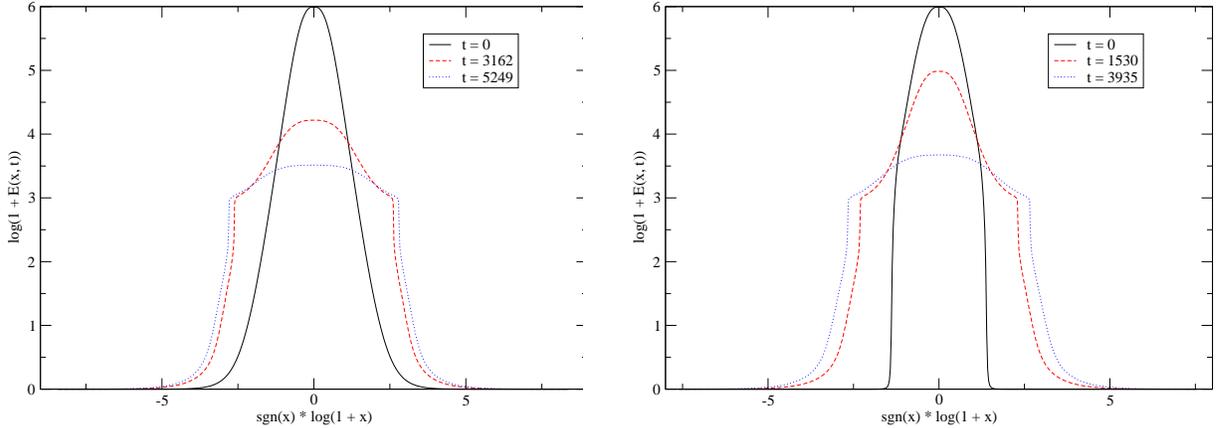

{\includegraphics[width=0.5\textwidth]{profiles1.eps}
~~~~~
\includegraphics[width=0.5\textwidth]{profiles2.eps}}
\caption{Different initial conditions both evolve towards a state that has a flat line of energy at $E_{\mathrm{F}}$ before decaying at the edges. In the special case where $\beta(E)$ is a pure power law, the observation of all initial conditions approaching the same limit is known to hold precisely. Logarithmic scales are used on both axes in order to make the important features visible.}
\label{profiles}
\end{figure}

\begin{figure}
\centering
\includegraphics[width=.7\textwidth]{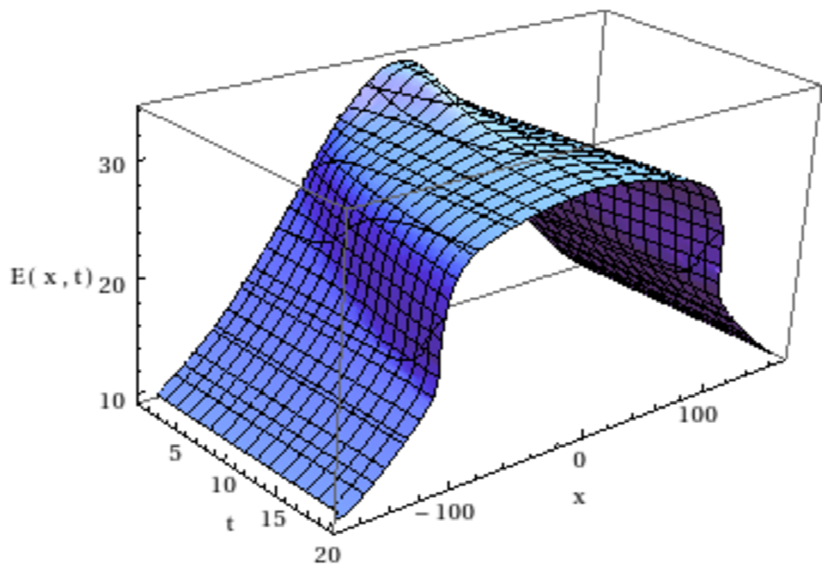}
  \caption{Evolution of an energy distribution for $\beta(E)$ similar to that in figure \ref{interpolated-beta}. The formation of a sharp boundary between high and low energy regions is a robust feature that appears for a wide range of initial conditions.}
\label{numeric}
\end{figure}

We present the results of a few of our simulations in figures \ref{profiles} and \ref{numeric}.

The simulations we have performed exhibit the following general behavior:
\begin{itemize}
\item
At early times, the evolution depends in detail on the initial condition. However the distributions generally approach one with an approximately uniform energy density $E_{\mathrm{F}}$ in a central region, separated by a sharp boundary from a region of low energy.
\item
The evolution from the initial energy distribution to an energy distribution with a plateau at $E \sim E_{\mathrm{F}}$ happens on a much shorter timescale than the subsequent decay of this flattened distribution.
\end{itemize}
These observations are consistent with the qualitative picture described in the previous section and summarized in figure \ref{intHag}. In particular, the last observation matches our expectations for holographic confining gauge theories (or their dual gravity theories) where the evolution of an initial energy distribution to a nearly-static plasma-ball (black hole) configuration is expected to occur much faster than the subsequent evaporation of the plasma ball (black hole).

The discussion in this section has been mainly qualitative, backed up by numerical simulations. In appendix B, we include a discussion of some rigorous analytic results pertaining to solutions of general equations of the form (\ref{Etwo2}). These may be useful in future investigations.

\section{Discussion}

In this note, we have illustrated by use of a simple stochastic toy model a connection between the density of states for subsystems of a system with a network of locally interacting degrees of freedom and the qualitative dynamics of energy distributions in the system. Depending on whether the density of states for sites in our model has a growth rate slower, faster, or equal to Hagedorn growth ($\rho(E) \sim e^{cE}$) we find diffusive, clustering, or static behavior for the energy distribution.

It is an intriguing possibility that our observations may help us understand why certain large $N$ gauge theories are able to describe the physics of black holes in their dual gravitational description. One of the unusual features that distinguishes large $N$ gauge theories from more typical field theories is the existence of Hagedorn and super-Hagedorn regimes in the density of states. We have seen, within the context of our toy model, a direct connection between these behaviors and trapping / clustering of energy. Thus, it is plausible that in the full large $N$ field theories with gravitational duals, the formation and long lifetime of black hole states may also be at least partly explained by the behavior of the density of states for subsystems of the field theory.

While our toy model appears to give results for the dynamics of localized energy distributions that are qualitatively similar to our expectations from certain field theories, we emphasize that this type of model is far too simplistic to give an accurate model of the field theory more generally. For example, while we have incorporated conservation of energy into our model, we have not included momentum as a conserved quantity. Thus, for example, the toy model cannot accurately model radiative transfer of energy. In the future, it may be interesting to consider more detailed models that incorporate additional features of the field theory to see whether we can find behavior matching a larger class of gravitational phenomena.

It is interesting to note that the macroscopic equations (generalizing (\ref{Etwo2})) for a more general stochastic model including conserved momentum would take a form similar to hydrodynamics equations. At least for near-equilibrium thermal states (and perhaps more generally for large $N$ theories) there is a general expectation that hydrodynamics should indeed provide an accurate description of the field theory dynamics. Furthermore, in recent work, specific hydrodynamical models associated with large $N$ gauge theories have been shown to have solutions in one-to-one correspondence with solutions of Einstein's equations expanded around black brane solutions in gravity with a negative cosmological constant \cite{Bhattacharyya:2008jc}. It would be interesting to see if one could reproduce these hydrodynamics equations as the macroscopic equations of a stochastic model for which the transition rates depend on both momentum and energy. In this case, one might be able to understand more directly how the
hydrodynamic coefficients (viscosity, \textit{etc}...) arise from microscopic properties of the underlying field theory. Additionally, such a construction would demonstrate that models only slightly more complicated that the one we considered here are capable of giving rise to detailed gravitational dynamics.

\subsection*{Acknowledgments} We thank B. Czech and J. McGreevy for helpful discussions and feedback on some of our results.  BGS is supported by a Simons Fellowship through Harvard University. This research is supported in part by the Natural Sciences and Engineering Research Council of Canada.

\appendix

\section{Example of diffusion with a Hagedorn regime}

In this section, we consider a simple example to see the timescale for diffusion of a localized collection of energy in a case where the ordinary diffusion equation governs diffusion for $E < E_{\mathrm{H}}$ while the diffusion constant drops to zero (as in a Hagedorn regime for our model) for $E>E_{\mathrm{H}}$. Thus, we would like to find the solution of
\[
\partial_t E = \partial_x^2 \Phi(E)
\]
where
\[
\Phi(E) = \left\{ \ba{ll}  E & E < E_{\mathrm{H}} \cr 0 & E \ge E_{\mathrm{H}} \ea \right.
\]
with initial conditions
\[
E(x,0) = \left\{ \ba{ll} 0 & |x| > a \cr E_0 & |x| < a \ea \right. \; ,
\]
where we assume that $E_0$ is greater than $E_{\mathrm{H}}$. We would like to compare our result to the simple diffusion case with $\Phi(E) = E$ in order to see the effects of having a Hagedorn regime.

As discussed in section 5, the solution for $t<T$ is expected to be discontinuous at some points $\pm R_{\mathrm{H}}(t)$, with $E = E_0$ for $|x|< R_{\mathrm{H}}(t)$ and $E$ decreasing from $E = E_{\mathrm{H}}$ as $|x|$ increases from $R_{\mathrm{H}}$. The point of discontinuity $R_{\mathrm{H}}(t)$ decreases with time from $x=a$, reaching $x=0$ at time $T$.

Suppose at some time we have $R_{\mathrm{H}}(t) = R$. Since the energy density jumps from $E_0$ down to $E_{\mathrm{H}}$ at $x=R_{\mathrm{H}}$, the rate of change of the energy in the region $0 \le x \le R$ at this instant will be $-(E_0 - E_{\mathrm{H}})\dot{R}_{\mathrm{H}}(t)$. This must equal the energy current just to the right of $R_{\mathrm{H}}$, so we must have
\be
\label{bc}
(E_0 - E_{\mathrm{H}})\dot{R}_{\mathrm{H}}(t) = \partial_x E (R_{\mathrm{H}}^+,t) \; .
\ee

Our strategy will be to find a solution $e(x,t)$ to the ordinary diffusion equation such that the point $R_{\mathrm{H}}(t)$ at which $e(R_{\mathrm{H}}(t),t) = E_{\mathrm{H}}$ satisfies the equation (\ref{bc}). In this case, the function
\[
E(x,t) = \left\{ \ba{ll} e(|x|,t) & |x| > R_{\mathrm{H}}(t) \cr E_0 & |x| < R_{\mathrm{H}}(t) \ea \right. \;
\]
will be a solution of our original differential equation, since it is static for $x<R_{\mathrm{H}}$, solves the ordinary diffusion equation for $x > R_{\mathrm{H}}$, and satisfies the boundary condition (\ref{bc}) by construction.

We will show that the desired function $e(x,t)$ is the solution to the ordinary diffusion equation with initial conditions
\[
e(x,0) = \left\{ \ba{ll} 0 & x > a \cr A & -\infty < x < a \ea \right. \; ,
\]
where $A$ is a constant determined by $E_0$ and $E_{\mathrm{H}}$. For these initial conditions, we have
\beas
e(x,t) &=& {A \over \sqrt{4 \pi t}} \int_{-\infty}^a \textup{d} \tilde{x} e^{-{(\tilde{x} - x)^2 \over 4t}} \cr
&=& {A \over 2} \left ( 1 + {\rm erf} \left( {a - x \over 2 \sqrt{t}} \right) \right )
\eeas
Now, choosing $A$ so that $E_{\mathrm{H}}$ is between $A/2$ and $A$, we can define $R_{\mathrm{H}}(t)$ as above by
\[
E_{\mathrm{H}} = e(R_{\mathrm{H}}(t),t) \; ,
\]
which gives
\[
R_{\mathrm{H}}(t) = a - 2 \sqrt{t} \;  {\rm erf}^{-1} \left ( 2 {E_{\mathrm{H}} \over A} - 1 \right ) \; .
\]
Now, using
\[
\partial_x e = -{A \over \sqrt{4 \pi t}} e^{-{(x-a)^2 \over 4 t}}
\]
we can check that
\[
{\partial_x e (R_{\mathrm{H}}(t),t) \over \dot{R}_{\mathrm{H}}(t)} = {A \over \sqrt{4 \pi}}{e^{-({\rm erf}^{-1}(2E_{\mathrm{H}}/A - 1))^2} \over {\rm erf}^{-1}(2E_{\mathrm{H}}/A - 1)} = {\rm constant} \; .
\]
Thus, we obtain the desired solution by choosing $A$ such that
\be
(E_0 - E_{\mathrm{H}}) = {A \over \sqrt{4 \pi}}{e^{-({\rm erf}^{-1}(2E_{\mathrm{H}}/A - 1))^2} \over {\rm erf}^{-1}(2E_{\mathrm{H}}/A - 1)} \; .
\label{Asol}
\ee
In summary, the solution is
\be
E(x,t) = \left\{ \ba{ll} {A \over 2} \left ( 1 + {\rm erf} \left( {a - x \over 2 \sqrt{t}} \right) \right ) & |x| > R_{\mathrm{H}}(t) \cr E_0 & |x| < R_{\mathrm{H}}(t) \ea \right. \;
\label{Hag}
\ee
with
\[
R_{\mathrm{H}}(t) = a - 2 \sqrt{t} \; {\rm erf}^{-1} \left ( 2 {E_{\mathrm{H}} \over A} - 1 \right )
\]
and $A$ given by (\ref{Asol}). More explicitly, $A$ is determined in terms of $E_{\mathrm{H}}$ and $E_0$ by
\[
A = E_{\mathrm{H}} { 2 \over 1 + {\rm erf}(I)}  \in [E_{\mathrm{H}}, 2 E_{\mathrm{H}}]
\]
where $I$ is the solution to
\[
\sqrt{\pi} I e^{I^2} (1 + {\rm erf}(I)) = {E_{\mathrm{H}} \over E_0 - E_{\mathrm{H}}} \; .
\]
For comparison, the solution of the ordinary diffusion equation with the same initial conditions is
\be
E_{\mathrm{diff}}(x,t) = {E_0 \over 2}\left({\rm erf} \left({x+a \over 2 \sqrt{t}}\right)-{\rm erf} \left({x-a \over 2 \sqrt{t}}\right)\right)
\label{noHag}
\ee

To see the quantitative effects of having a Hagedorn regime for $E > E_{\mathrm{H}}$, we can compare the flow of energy outside the region $[-a,a]$ for the solutions $(\ref{Hag})$ and $(\ref{noHag})$. In each case, we calculate the energy in the region $|x|>a$ as a function of time. With the Hagedorn regime, we find that while $R_{\mathrm{H}}(t)>0$ (i.e. while the energy density at $x=0$ is still greater than $E_{\mathrm{H}}$),
\beas
E_{|x|>a} &=& A \int_a^\infty \textup{d}x \; \left(1 + {\rm erf}\left({a - x \over 2 \sqrt{t}}\right)\right) \cr
&=& 2 A \sqrt{t \over \pi} \cr
&\approx& 4 E_{\mathrm{H}} \sqrt{t \over \pi} \qquad \qquad (E_0 \gg E_{\mathrm{H}})
\eeas
For the simple diffusion case, we find
\beas
E_{|x|>a} &=& E_0 \int_a^\infty \textup{d}x \; \left({\rm erf}\left({a+x \over 2 \sqrt{t}}\right) + {\rm erf}\left({a-x \over 2 \sqrt{t}}\right) \right) \cr
&=& 2 E_0 \sqrt{t \over \pi}(1 - e^{-{a^2 \over t}}) + 2 E_0 a \left(1 - {\rm erf}\left({a \over  \sqrt{t}}\right) \right) \cr
&\approx& 2 E_0 \sqrt{t \over \pi} \qquad \qquad (t \ll a^2)
\eeas
Thus, at early times, the energy flow away from the central region is smaller by a factor $2 E_{\mathrm{H}} / E_0$ with a Hagedorn regime.

We note that the time before the central maximum in the energy distribution decreases to the Hagedorn threshold $E_{\mathrm{H}}$ (equal to the time $T$ when $R_{\mathrm{H}}(T)=0$) is
\be
\label{Hagtime}
T = {1 \over 4}\left({a \over {\rm erf}^{-1}(2E_{\mathrm{H}}/A - 1)}\right)^2 \approx {\pi \over 4} \left( a {E_0 \over E_{\mathrm{H}}} \right)^2 \qquad \qquad E_0 \gg E_{\mathrm{H}} \; .
\ee
In the case of simple diffusion, the time when $E(x=0) = E_{\mathrm{H}}$ is
\be
\label{noHagtime}
T_{\mathrm{diff}} = \left({a \over 2 {\rm erf}^{-1} \left({E_{\mathrm{H}} / E_0}\right)} \right)^2 \approx {1 \over \pi} \left( a {E_0 \over E_{\mathrm{H}}} \right)^2 \; .
\ee
This differs from (\ref{Hagtime}) only by a factor of order 1, so over longer time scales, the flow rate of energy out of the central region becomes comparable for the two cases. The reason is that as energy leaves the central region, it does not escape quickly to infinity as it would in a relativistic field theory, but rather diffuses slowly away. This results in a buildup of energy outside the central region which results in a smaller $\partial_x E$ at the boundaries of the central region, and thus a smaller flow rate. Thus, at these later times, the flow rate away from the central region is limited by the speed at which energy can diffuse away. This is the same in both cases, so the effects of the Hagedorn regime in $\Phi(E)$ is masked somewhat.

\section{Concentration Comparison}\label{appA}

\newtheorem*{theo}{Concentration comaprison}
The equation
\be
\partial_t E = \partial_x^2 \Phi(E) \; ,
\label{DE}
\ee
equivalent to (\ref{Etwo2}), is commonly referred to as the Filtration equation and appears in the study of
diffusion in porous media \cite{vazquezbook}. In this appendix, we point out a few general results pertaining to solutions of this equation that may be useful in future investigations.

For smooth axisymmetric energy distributions that monotonically decrease in the radial direction, it follows from (\ref{Etwo2}) that that the maximum energy density at $r=0$ decreases monotonically with time.\footnote{This is a consequence of the so-called maximum principle.}
The qualitative behaviour is that energy diffuses from large densities at small radii, where $\nabla^2 E<0$,
to small densities at large $r$, where $\nabla^2 E>0$. Note that (each component of) $\nabla E$ stays non-positive at all $r\neq 0$;
therefore $r=0$ is the only local maximum at all times.\footnote{For $\nabla E$ to become negative at $r\neq 0$
 it has to first vanish. The point where $\nabla E|_{r\neq0}$ first vanishes is an inflection point, i.e. the components of $\nabla \nabla^2 E$ are negative.
 Then, taking the gradient of the  (\ref{EtwoA}) shows that $\nabla E$ remains non-positive:
\bea
\p_t \nabla E=-\tilde{\beta}'\nabla^3E<0.\nn
\eea
}

Consider a ball $B(R)$ of radius $R$ centered at $r=0$. We denote by $M_R(t)$ the total amount of mass inside the ball:
\bea
M_R(t)=\int_{B(R)} \mathrm{d}^dx E(t)\nn.
\eea
This mass decreases as a function of time as:
\bea\label{derM}
\p_t M_R=-\tilde{\beta}'\int_{S(R)}\nabla E(t)<0,
\eea
where we have integrated (\ref{Etwo2}) and $S(R)$ is the boundary of $B(R)$. We can also define the diffusion time $t_*(R,\epsilon)$ to be the shortest time it takes for $M_R(t)$ to drop to $\epsilon$ times
the total mass, i.e. $M_R(t_*)=\epsilon M_\infty(0)$.

In order to obtain quantitative bounds on the time scales for evolution of energy distributions for general $\Phi(E)$, we can use  the following comparison theorem \cite{vazquez96,vazquezbook}:
  \begin{theo}
  Let $e_i(x,t)$, $i=1,2,3$ be solutions to $\p_te_i=\nabla^2\Phi_i(e_i)$, respectively, for monotonically increasing $\Phi_i(e_i)$ and $\Phi_i(0)=0$.
  If $e_1(x,0)=e_2(x,0)=e_3(x,0)$ and $\Phi_1(e)\succeq \Phi_2(e)\succeq \Phi_3(e)$ for all $e$\footnote{$f(e)\succeq g(e)$ if and only if $\int_0^e f(e')de'\geq \int_0^e g(e')de'$.} then, $M_1(R,t)\leq M_2(R,t)\leq M_3(R,t)$ for all $R$ and $t$, where
  $M_i(R)=\int_{B(R)}\mathrm{d}^dx\:e_i$. Therefore:
  $t_*^1(R,\epsilon)\leq t_*^2(R,\epsilon)\leq t_*^3(R,\epsilon)$.
     \end{theo}

As an example, we consider the evolution of an energy distribution in our model equation (\ref{Etwo2}) in the case where $\beta(E)$ has Hagedorn behavior $\beta'(E)=0$ for $E > E_{\mathrm{H}}$ and typical field theory behavior $\beta(E) = E^\alpha$ with $0 < \alpha < 1$ for $E < E_{\mathrm{H}}$. In regions where the energy is less then $E_{\mathrm{H}}$, the evolution equation (\ref{Etwo2}) reduces to the well-known super-fast diffusion equation:
\bea\label{fast}
\p_t E= \alpha(1-\alpha) \nabla \cdot (E^{\alpha-2}\nabla E) \; .
\eea
 The super-fast diffusion takes its name from its singular behavior at $E\to 0$. From (\ref{current}), we see that the energy current is
 \bea\label{velocity}
J_E=- \alpha(1-\alpha) E^{\alpha-2}\nabla E\nn,
 \eea
which diverges for $E\to 0$. This singular behavior means that energy can escape to infinity in finite time. However, in realistic models, the density of states will be modified from the $\rho(E) = \exp(A E^\alpha)$ behavior at the lowest energies so that this behavior is avoided. Here, we will avoid the problematic low-energy behavior by considering initial Cauchy data with a small background energy density,
\bea\label{ic}
E(x,t=0)=e(x,t=0)+E_{\mathrm{B}}\nn,
\eea
where $e(x,t)$ is a non-negative function that vanishes at spatial infinity. Therefore, the problem we address here is the relaxation of far-from-equilibrium initial distributions evolving on a uniform background. Note that introducing a background, no matter how small, dramatically changes the solutions to the fast-diffusion equation since it cuts off the propagation speed of perturbations.\footnote{In $d$ dimensions there are  no solutions to (\ref{fast}) for $\alpha\leq d/2$ due to the phenomenon of instantaneous extinction \cite{vazquez96}. However, in the presence of the background, the solutions exist at least for all $0<\alpha$.}

In summary, we wish to consider a solution to the equation
\be
\partial_t E = \partial_x^2 \Phi(E) \; ,
\label{filtration}
\ee
with
\[
\Phi(E) = \left\{ \ba{ll}  -\alpha E^{\alpha -1} & E < E_{\mathrm{H}} \cr -\alpha E_{\mathrm{H}}^{\alpha - 1} & E \ge E_{\mathrm{H}} \ea \right.
\]
with initial conditions
\[
E(x,0) = \left\{ \ba{ll} E_{\mathrm{B}} & |x| > a \cr E_0 & |x| < a \ea \right. \; ,
\]
where we assume that $E_0>E_{\mathrm{H}}>E_{\mathrm{B}}$.

If $E(x,t)$ is the desired solution, then we note that $e(x,t) = E(x,t) - E_{\mathrm{B}}$ will be a solution to the equation (\ref{filtration}) with $\Phi(e) = \Phi(e + E_{\mathrm{B}}) + C$, where $C$ is an arbitrary constant. In order to apply the concentration comparison theorem, we will choose $C$ so that $\Phi(0) = 0$. Thus, we consider the equation (\ref{filtration}) with
\be
\Phi(e) = \left\{ \ba{ll} \alpha E_{\mathrm{B}}^{\alpha-1} - \alpha (e + E_{\mathrm{B}})^{\alpha - 1}  & e < E_{\mathrm{H}} - E_{\mathrm{B}} \cr \alpha E_{\mathrm{B}}^{\alpha -1} - \alpha E_{\mathrm{H}}^{\alpha - 1} & E \ge E_{\mathrm{H}} - E_{\mathrm{B}} \ea \right.
\label{phicc}
\ee
and
\[
e(x,0) = \left\{ \ba{ll} 0 & |x| > a \cr E_0 - E_{\mathrm{B}} & |x| < a \ea \right. \; ,
\]
We would like to know how long it takes before the energy density is everywhere smaller than the Hagedorn density. To obtain bounds on this time scale, we use the concentration comparison theorem, choosing $\Phi_2(e)$ as in (\ref{phicc}), and taking $\Phi_1$ and $\Phi_3$ to be of the form
\[
\Phi_i(E) = \left\{ \ba{ll}  B_i E & E < E^M_i \cr 0 & E \ge E^M_i \ea \right.
\]
for which we obtained the solution in the previous section. By choosing $B_i$ and $E^M_i$ optimally so that the conditions of the theorem are satisfied\footnote{Specifically, the optimal bounds come by ensuring that $\int_0^{E_0-E_{\mathrm{B}}} \Phi_1 = \int_0^{E_0-E_{\mathrm{B}}} \Phi_2 =\int_0^{E_0-E_{\mathrm{B}}} \Phi_3$ and then choosing $B_1$ and $B_3$ as large and as small as possible so that $\int_0^E \Phi_1 \le \int_0^E \Phi_2 \le \int_0^E \Phi_3$ for all $E<E_0 - E_{\mathrm{B}}$.}, we can show that the time before the energy density is less than $E_{\mathrm{H}}$ everywhere is bounded by
\[
{\pi \over 8 (1- \alpha) E_{\mathrm{H}}^\alpha} (a E_0)^2  \le T \le {\pi (1-\alpha) \over 4 \alpha  E_{\mathrm{B}}^\alpha} (a E_0)^2 \; .
\]

\end{document}